# Titanium Contacts to Graphene: Process-Induced Variability in Electronic and Thermal Transport

*Keren M. Freedy[1], Ashutosh Giri[2], Brian M. Foley[2], Matthew R. Barone[1], Patrick E. Hopkins[1,2], Stephen McDonnell[1]*

[1]Department of Materials Science and Engineering, University of Virginia, Charlottesville, Virginia, 22904, United States

[2]Department of Mechanical and Aerospace Engineering University of Virginia, Charlottesville, Virginia, 22904, United States

CORRESPONDING AUTHOR: mcdonnell@virginia.edu.

ABSTRACT

Contact Resistance ($R_C$) is a major limiting factor in the performance of graphene devices. $R_C$ is sensitive to the quality of the interface and the composition of the contact, which are affected by the graphene transfer process and contact deposition conditions. In this work, a linear correlation is observed between the composition of Ti contacts, characterized by X-ray photoelectron spectroscopy, and the Ti/graphene (Gr) contact resistance measured by the transfer length method. We find that contact composition is tunable via deposition rate and base pressure. Reactor base pressure is found to effect the resultant contact resistance. The effect of contact deposition conditions on thermal transport measured by time-domain thermoreflectance is also reported and interfaces with higher oxide composition appear to result in a lower thermal boundary conductance. Possible origins of this thermal boundary conductance change with oxide composition are discussed.

INTRODUCTION

One of the major challenges associated with the design of two dimensional (2D) devices is the large contact resistance ($R_C$) at the interface between the 2D material and the metal.[1, 2] The contribution of $R_C$ to the total device resistance increases as the channel length is scaled down, meaning that $R_C$ can ultimately be the limiting factor in the performance of 2D devices[3-5] Graphene has excellent electrical and thermal transport properties,[6, 7] making it particularly important to understand the mechanisms of losses across the meta/Gr interface. A large volume of recent experimental[4, 8-13] and theoretical[14-16] work is focused on understanding the chemistry and electronic properties of the metal/Gr interface. Many assume ideal interfaces in which $R_C$ is explained by intrinsic interactions between graphene and the metal such as effects of orbital hybridization, electrochemical equalization, and other mechanisms which cause changes in the electronic structure of graphene due to the presence of a metal overlayer.[5, 14] The effects of processing conditions on the chemistry and properties of the contact are often overlooked. The details of graphene processing procedures and contact deposition conditions such as base pressure and deposition rate are rarely reported in device studies, even in those which focus specifically on characterization of contacts.[4, 9, 10, 17, 18] Several works demonstrate that $R_C$ is independent of the gate voltage and the number of graphene layers, indicating that it is dominated by the properties of the metal/Gr interface.[9, 12, 19] This warrants more thorough interface characterization. Titanium was selected for this work as it is commonly used as a contact or adhesion layer for graphene due to its low work function and low electron Schottky barrier.[10, 18, 19]

EXPERIMENTAL DETAILS

To fabricate samples for this experiment, commercial graphene grown by chemical vapor deposition (CVD) on Cu foil was transferred to $SiO_2$ by a polymethyl methacrylate (PMMA) carrier film. [20, 21] A solution of 30 mg/mL PMMA dissolved in chlorobenzene was spin-coated at 4000 rpm for 30 seconds onto the Gr/Cu stack. The PMMA/Gr/Cu stack was cured at 60 °C for 10 minutes. The stack was placed in 3:1 deionized (DI) $H_2O$:$HNO_3$ for 1 minute followed by DI $H_2O$ for 1 minute to remove graphene from the back of the foil. This was repeated twice. The Cu foil was then dissolved in a solution of 0.5 M ammonium persulfate (APS) for a total of 21 hours. The PMMA/graphene film was then transferred onto a 300 nm $SiO_2$/Si wafer. Before transfer, the wafer was cleaned with methanol, acetone, and DI water. The Gr/$SiO_2$ was left to air dry for 30 minutes and was then heated to 180 °C for 5 minutes. Following this process, PMMA was dissolved in acetone. The samples were then annealed in ultra-high vacuum at 350-410 °C for three hours to remove PMMA residues.

A 5 nm film of titanium was then deposited onto Gr/$SiO_2$ in a HV electron beam evaporator at pressures of $10^{-7}$-$10^{-6}$ Torr and deposition rates ranging from 0.01 to 0.5 nm/s, indicated by a quartz crystal monitor. Samples for TLM measurements were fixed with a shadow mask described elsewhere.[13] The impact of resist residues on contact resistance will be the focus of future work. The samples were not exposed to atmosphere immediately following the deposition of Ti. In other words, Au was deposited to cap the samples prior to removal from UHV in order to prevent further oxidation of the Ti layer upon air exposure. Au films of 500 nm, 80 nm, and 2 nm were deposited on samples for TLM, thermal measurements, and XPS, respectively. X-ray photoelectron spectroscopy data was collected with a monochromated X-ray source at a pass energy of 50 eV in a UHV system described previously.[22] Spectra were deconvoluted using kolXPD software[23] to extract relative compositions of Ti metal and Ti oxide.

TLM data was acquired using 19 micron gold-plated tungsten probe tips (CascadeMicrotech, 154-001) in a probe station (JmicroTechnology, LMS-2709) connected to a SourceMeter unit (SMU, Keithley Instruments 2612A) with an applied source current of 1 mA. The data was acquired under ambient conditions within 12 to 14 days of the initial graphene transfer and within one week of contact deposition. Prior to measurement, the samples were stored in a desiccator. On each sample, sixteen TLM structures were measured and the resistances corresponding to each contact separation distance were averaged. Results acquired on the same samples after six months of air exposure show a trend consistent with the original analysis. The contact separation distances for this TLM structure are 47.3±1.3, 71.9±1.2, 97.9±2.0, 122.5±1.7 and 147.4±1.0 μm with a contact size of 200x450 μm$^2$.[13] The contact resistance of the interface was determined by plotting average contact resistance vs. separation distance and extrapolating the y- intercept from a linear fit to the data. Other graphene TLM studies utilize a wide range of TLM geometries typically processed by photolithography with contact spacings less than 100 μm.[9, 10, 24, 25] Typical reactor base pressures and deposition rates were not report and no detailed comparison with these reports in possible. The present work reports on the effect of metal deposition conditions on contact resistance and the minimization of potential variations induced by photoresist residue was therefore avoided by using a shadow mask.

Thermal boundary conductance was measured using time-domain thermoreflectance (TDTR). Laser pulses emanate from a Ti:Sapphire oscillator with an 80 MHz repetition rate, which are energetically split into a pump path (that provides the heating event for the sample) and probe path (that is time-delayed in reference to the pump pulses) that is used to monitor the thermoreflectance of the sample under consideration as a function of pump-probe time delay. The pump path is modulated at 10 MHz and a lock-in amplifier is utilized to monitor the ratio of the in-phase to out-of-phase signal of the reflected probe beam ($-V_{in}/V_{out}$) at the pump modulation frequency for a total of 5.5 ns after the initial heating event. Several TDTR scans are performed at different locations across the samples to ensure repeatability of the measurements, and the data are fit with a model that accounts for thermal diffusion in a two layer system by fitting for $h_K$ across the Au/SiO$_2$ interface.

The value of $h_K$ provides a quantitative metric for the efficacy with which energy is exchanged across interfaces.[26] Note, in practice these reported values represent the thermal boundary conductance across an Au/SiO$_2$ contact with contributions from the Ti and graphene layers and contaminant interfaces. These measured Au/SiO$_2$ thermal boundary conductance values represent a lumped conductance value that accounts for heat flow from the Au, across the Au/Ti interface, through the Ti layer, across the Ti/Gr interface, and finally across the Gr/SiO$_2$ interface. Due to the relatively small thicknesses of the Ti and graphene, this Ti/Gr layer is treated as the interfacial layer between the Au and SiO$_2$, and thus these values for $h_K$ are indicative of the thermal conductance across an Au/SiO$_2$ contact with Ti/Gr in between, consistent with prior TDTR analyses and descriptions on similar systems.[13, 27] The appropriate analysis procedure to measure $h_K$ and the details of the experimental setup are given elsewhere.[28-30] The specific assumptions in our analysis regarding similar Au/Ti/Gr/SiO$_2$ systems are outlined in detail in our previous work.[13]

RESULTS

We have found that oxide composition is largely dependent on the contact deposition conditions. Titanium is highly reactive and will readily oxidize under high-vacuum deposition conditions. As others have suggested,[19, 27, 31-34] the adsorption of oxidizing species onto the substrate surface during deposition will affect the chemistry of the contact which is expected to manifest in the electrical and thermal properties of the interface. Figure 1 shows oxide composition vs. deposition rate for samples fabricated from three individually transferred pieces of graphene.

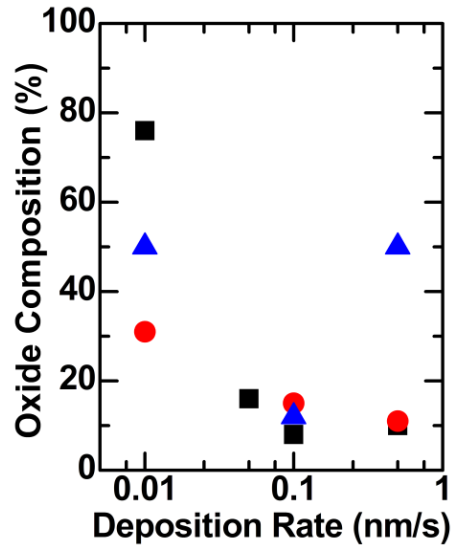

**Figure 1.** Plot of Ti oxide composition vs. deposition rate at a pressure of 1x10$^{-7}$ Torr on Gr/SiO$_2$ samples. Each identical marker shape represents samples cut from the same piece of graphene.

Each color represents a single piece of graphene transferred to $SiO_2$ and subsequently split into three (or four) samples to receive metal deposition at three (or four) different deposition rates. Sample-to-sample variability is observed, but there appears to be a trend of decreasing oxide composition with increasing deposition rate. The deposition rate determines the impingement rate of Ti atoms on the surface of the substrate relative to the impingement rate of the oxidizing species from residual gases. It is therefore expected that higher deposition rates result in lower oxide composition, since at higher deposition rates, Ti atoms arrive at the sample surface at faster rates than oxidizing species in the chamber. Anomalous data points can be explained by the presence of additional oxidizing species from PMMA residues which will be addressed in the Discussion section.

Base pressure also has a substantial effect which can dominate over deposition rate. The base pressure is a measure of the quantity of residual gases in the chamber. Depositing at higher pressures increases the amount of oxidizing species available for reaction with Ti, and depositing at lower deposition rates increases the fraction of Ti atoms which will react with oxidizing species upon reaching the surface. This is observed in Figure 2. To overcome the issue of sample-to-sample variability, each sample represented in Figure 2 was cut from a single piece $Gr/SiO_2$ produced in a single transfer. Two out of the three samples were deposited on at the same rate and different base pressures, and two out of three were deposited on at the same base pressure but different rates. In Figure 2(a), (i) corresponds to a deposition $1 \times 10^{-7}$ Torr and a rate of 0.01 nm/s, (ii) corresponds to a deposition at $1 \times 10^{-7}$ Torr and a rate of 0.1 nm/s and (iii) corresponds to a pressure girof $1 \times 10^{-6}$ Torr at a rate of 0.1 nm/s. The corresponding TLM data for each are shown in Figure 2(b). The results indicate that base pressure has a stronger effect on contact composition than deposition rate, since (iii) shows a comparable oxide composition of 78% compared with (i) which is 67% and yet shows markedly different $R_C$. It is known that UHV depositions result in cleaner interfaces and improved $R_C$ for unreactive metals like Au.[3] Our comparison of samples processed under lower and higher base pressures show that $R_C$ might be dominated by the composition of the interface rather than the composition of the contact itself, as contacts with similar oxide compositions ((i) and (iii)) exhibit a large difference in $R_C$. The larger error bars and confidence interval in the TLM data for (iii) are also indicative of greater variability in measured $R_C$ throughout different regions of the sample. Comparison of spectra (i) and (ii) in Figure 2 illustrate the results reported in Figure 1 concerning the effect of deposition rate at a constant base pressure.

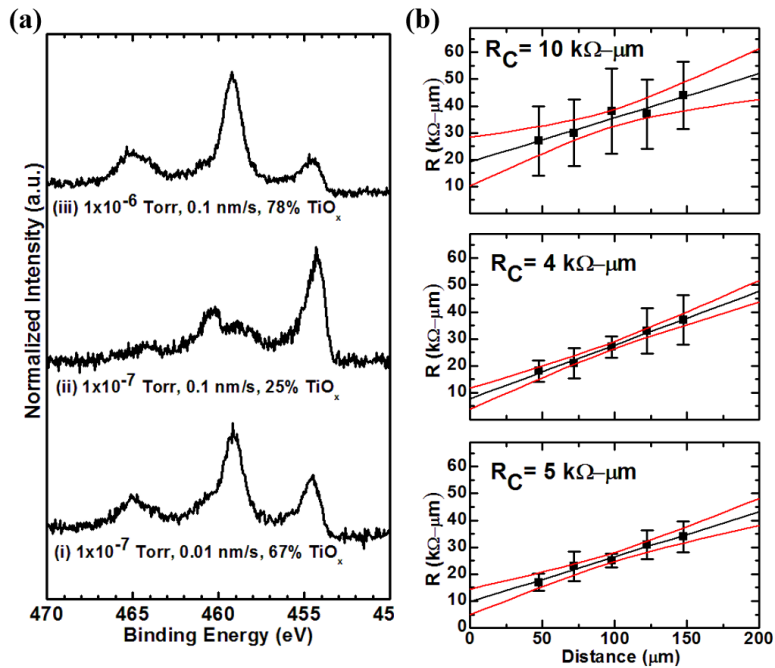

**Figure 2.** (a) Ti 2*p* core-level spectra for Ti deposited onto samples cut from a single Gr/SiO$_2$ sample at different deposition conditions resulting in different oxide compositions. (b) Corresponding TLM results for each sample where black line represents the linear fit and the red lines represent the upper and lower 95% confidence bounds

We have observed an overall correlation between the oxide composition of the contacts shown in Figure 3. While large sample-to-sample variability is observed, the data has a linear correlation coefficient of 0.7. The linear correlation coefficient describes the extent to which two variables support a linear relation.[35] Thus, a value of this linear correlation coefficient close to approaching unity indicates a linear relationship likely exists where the probability of correlation depends on the number of data points acquired. For the thirteen values reported in this work, the probability of a linear correlation is 99.2%. Thus we conclude there exists a linear relationship between oxide composition and $R_C$. Differences in the cleanliness of the interface observed in Figure 2 might also explain why contacts of similar oxide composition show large variation in $R_C$ as seen in Figure 3.

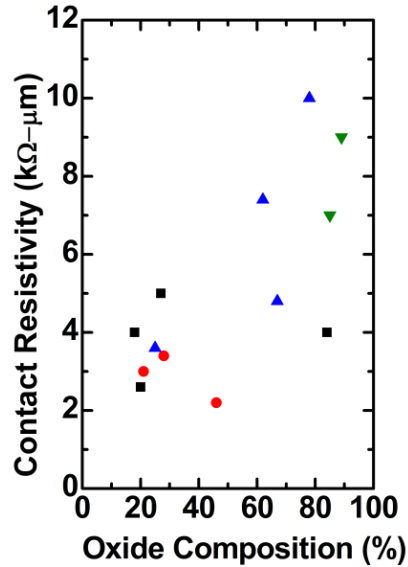

**Figure 3.** Plot of width-normalized contact resistivity as a function of oxide composition showing a linear trend with a correlation coefficient of 0.7. Each set of identical markers on the plot corresponds to samples which were cut from the same piece of transferred graphene but processed under different conditions

The effects of contact processing conditions manifest in thermal transport properties. Figure 4(a) shows XPS spectra acquired for three samples fabricated with three different deposition rates and Figure 4(b) shows the corresponding TDTR data as a function of oxide composition. XPS shows significant oxide composition at the slowest deposition rate of 0.01 nm/s. The oxide composition decreases between 0.01 and 0.05 nm/s. The thermal data indicates that thermal boundary conductance $h_K$ is inversely related to the oxide composition. For the deposition rate of 0.1 nm/s which resulted in the lowest oxide composition, $h_K$ =65±7 MW m$^{-2}$ K$^{-1}$, whereas for the slowest deposition rate which resulted in the highest oxide composition, $h_K$ =32±3 MW m$^{-2}$ K$^{-1}$ for the Au/SiO$_2$ interface where the effective interfacial regions between Au and SiO$_2$ for this analysis is the Ti/Gr layers, as mentioned previously.

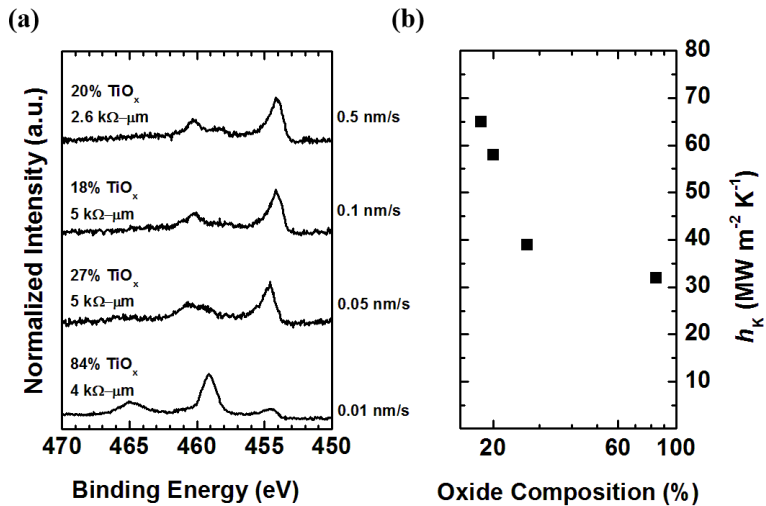

**Figure 4**. (a) Ti 2$p$ core-level spectra for Gr/SiO$_2$ deposited at different rates at a base pressure of 1x10$^{-7}$ Torr. (b) Time-domain thermoreflectance data for the same samples as a function of oxide composition.

The measured value of $h_K$ for the slower deposition rates matches very well with those measured for a similar Au/Ti/Gr/SiO$_2$ interface deposited at 0.05 nm/s and reported by Koh *et al.*[27] The twofold increase in $h_K$ with the faster deposition rate corresponds to the relative decrease in the oxide composition between the different deposition rates as shown in Figure 4(a). Thus, a higher oxide composition in the Ti layer at an Au/Ti/Gr/SiO$_2$ contact leads to a lower $h_K$ (higher resistance) than a lower oxide composition. Stated differently, our results suggest that to minimize the thermal resistance at an Au/Ti/Gr/SiO$_2$ contact, the Ti should be as metallic as possible. In contrast to thermal transport, electrical transport does not appear to be as sensitive to the composition of the contact for this particular sample, however the results shown in Figure 3 indicate that the reactor base pressure does have an impact on $R_C$.

DISCUSSION

It is apparent in Figure 1 that samples processed identically might result in different oxide compositions. A major source of variability in the Gr/metal interface chemistry is related to PMMA residue from the transfer process.[20, 36, 37] PMMA is typically removed by dissolution in acetone followed by an anneal in UHV at a temperature high enough to dissociate the various hydrocarbon species.[38] The thermal decomposition of PMMA is inherently a random process, and generated radicals can react with defects in the graphene or form longer polymer chains that cannot be removed.[39] Therefore, samples which undergo the same PMMA removal process can be left with different quantities of PMMA residue, and the quantity of PMMA residue is unlikely to be uniform across a single sample. Lee *et al.* have shown that a PMMA-free transfer process results in lower contact resistance than that which uses PMMA.[40] PMMA residues are known to dope graphene and alter its electronic properties.[38] Furthermore, transport across the Ti/Gr interface will be inhibited by the presence of contaminants which scatter charge carriers and obstruct hybridization between the graphene π-orbitals and Ti metal d-orbitals.[5, 14] Orbital hybridization will be inhibited both by the presence of polymer residues at the interface and by the presence of an oxidized contact rather than a metallic one. It has been shown that PMMA residues react with Ti overlayers.[41] Other sources of variability in both the interface and contact chemistry could be related to intrinsic defects in the CVD-grown graphene film or due to other extrinsic effects of transfer process including residual Cu, incomplete removal of graphene from the back of the Cu foil, wrinkles and tears in the film, or adsorbates. While measures can be taken to assess the quality and uniformity of the transferred graphene prior to device fabrication, such as characterization with Raman spectroscopy,[37, 38, 42] these defects are inherent to the transfer process and are fundamentally uncontrollable.

Despite the inevitable sample-to-sample variability, our results suggest that some degree of control over contact composition is achievable during the deposition process, particularly via deposition rate and base pressure. The overall linear correlation between oxide composition and $R_C$ summarized in Figure 3 is not surprising given that the electrical resistivity of $TiO_2$ is orders of magnitude higher than that of metallic Ti.[43] Since the contact resistance includes any contribution to resistance that is independent of the channel length[9], resistances within the Au and Ti layers, and at the Au/TiOx and graphene/TiOx interfaces are all contributors to the measured value. The results presented in Figure 3 indicate that the cleanliness of the graphene/TiOx interface likely dominates $R_C$ to a greater extent than the oxide composition.

The possible origins of the change in thermal boundary conductance with change in oxygen content of the Ti layer between the Au and graphene could manifest from various changes in electronic and vibrational scattering and interfacial transport in each layer of the Au/Ti/Gr/$SiO_2$ boundary region. The major contributors to this change in thermal resistance, $\Delta R = \Delta(1/h_K)$ ~15.9 $m^2$ K $GW^{-1}$, could be the change in $h_K$ at the Au/Ti interface, the change in thermal conductivity of the Ti, and the change in $h_K$ across the Ti/Gr/SiO$_2$ interface that would occur with a change in oxygen content in the Ti layer. In the extreme cases for thermal conductivity of the Ti layer, one would expect the Ti layer to either be fully TiO$_2$ or metallic Ti (clearly these are not the cases in our study, but we pose these extreme cases for exemplary purposes). Assuming the extreme cases of the thermal conductivity of amorphous TiO$_2$ (~1.2 W $m^{-1}$ $K^{-1}$)[44] and that of crystalline metallic Ti (~21.9 W $m^{-1}$ $K^{-1}$)[45], this leads to a change in resistance of this layer as $\Delta R = 5\times10^{-9}/1.2 - 5\times10^{-9}/21.9 \sim 4$ $m^2$ K $GW^{-1}$. We note this example calculation considers the extreme case to calculate the maximum hypothetical resistance change of this layer. As is evident, the above calculation for $\Delta R$ cannot explain the entire observed change in thermal boundary resistance with different oxygen content in the Ti layer (as previously mentioned, $\Delta R$ ~15.9 $m^2$ K $GW^{-1}$). We note also that size effects were not considered in this $\Delta R$ calculation.[46-48] Thus, the change in thermal resistance of the Ti layer cannot solely explain measured change in thermal boundary conductance.

Another possibility for the observed change in thermal conductance across the Au/Ti/Gr/SiO$_2$ region is the change in the Ti/Gr/SiO$_2$ thermal boundary conductance. Our previous work has demonstrated that changes in graphene surface chemistry induced from plasma functionalization (including oxygen functionalization) can lead to appreciable changes in thermal boundary conductance.[13, 49, 50] Thus, one could hypothesize that the change in oxygen stoichiometry in the Ti layer would also lead to changes in how the Ti reacts with residues and thereby lead to changes in the chemistry at the Ti/Gr interface; thus impacting thermal boundary conductance. We note that residual PMMA residue is present on all samples. Therefore it is presumed that all Gr/Ti interfaces will actually be TiO$_x$/Gr with some variation in amount of hydrocarbon incorporated.[41] At this time the impact of the variations in hydrocarbon incorporation at the interface on the thermal boundary conductance is unknown. Therefore, we cannot rule this out as a potential mechanism, and thus leave an intricate study of the chemistry effects on Ti/Gr thermal boundary conductance to future work.

Finally, we consider the change in thermal boundary conductance at the Au/Ti interface as a possible contributor to the measured change in $\Delta R$ of the Au/Ti/Gr/SiO$_2$ interfacial region. At pure metal/metal interfaces, the thermal boundary conductance is driven by the electron densities of states at the Fermi energies of the metals[51-54], and this corresponding thermal boundary conductance can be more than an order of magnitude greater than those at metal/non-metal interfaces.[55] While the thermal boundary conductances across Au/metal Ti and Au/TiO$_2$ interfaces have not been explicitly and reliably measured previously due to the exceptionally high $h_K$ affiliated with metal/metal interfaces, we can assume that the resistance associated with the metallic phase of Ti in contact with the Au will offer negligible resistance as compared to the non-metal oxide phases in the Ti layer. Indeed, typical values for thermal boundary conductances across Au/non-metal interfaces range from ~50 – 100 MW m$^{-2}$ K$^{-1}$, [53, 56-58] limited by the relatively narrow spectral phonon bandwidth in the Au. This corresponds to a $\Delta R$ of ~10 – 20 m$^2$ K GW$^{-1}$, on the order of our measured change in thermal boundary conductance with changes in oxygen content in the Ti ($\Delta R$ ~15.9 m$^2$ K GW$^{-1}$).

We estimate these various electron-electron and phonon-phonon resistances at the Au/Ti layer interface in more quantitative detail through the use of diffuse mismatch models (DMM). As mentioned previously, the electron DMM (eDMM) predicts the thermal boundary conductance between two materials with large electron densities of states compared to phonon density of states (e.g., at metal/metal interfaces).[51, 53, 54] Assuming values for the electron density of states at the Fermi energy and calculated Fermi velocities for Au and Ti[59, 60], we predict a thermal boundary resistance of $R_{ee,Au/Ti}$ = 0.17 m$^2$ K GW$^{-1}$ (thermal boundary conductance of 5,970 MW m$^{-2}$ K$^{-1}$ between the electronic systems in Au and Ti, assuming both are pure metals). This eDMM calculation thus predicts the thermal transport across the Au/Ti interface in the case when Ti is fully metallic. When the Ti layer is oxidized, this electron-electron interfacial thermal transport pathway will be reduced, and thus the Au phonon/Ti phonon interfacial thermal transport pathway can become a dominant conductance, since the electronic densities of states of the TiO$_x$ regions of the adhesion layer will be greatly reduced compared to the metallic Ti regions. Thus, we quantify this phonon-phonon thermal boundary resistance using the traditionally implemented phonon DMM (pDMM).[26] We calculate the phonon-phonon thermal boundary resistance of two cases: Au/Ti and Au/TiO$_2$ (rutile). In our pDMM calculations, we assume sine-type phonon dispersions of the longitudinal and two degenerate transverse acoustic modes in each material with zone edge phonon frequencies taken from Ref. [61] for Au (Γ→X direction), Ref. [62] for Ti (Γ→A direction), and Ref. [63] for rutile (Γ→A direction). From this, we predict phonon-phonon thermal boundary resistances of $R_{pp,Au/Ti}$ = 6.17 m$^2$ K GW$^{-1}$ ($h_{K,pp,Au/Ti}$ = 162 MW m$^{-2}$ K$^{-1}$) and $R_{pp,Au/TiO2}$ = 6.76 m$^2$ K GW$^{-1}$ ($h_{K,pp,Au/TiO2}$ = 148 MW m$^{-2}$ K$^{-1}$) for the Au/Ti and Au/TiO$_2$, respectively. Based on these eDMM and pDMM calculations, the predicted change in thermal boundary resistance associated with the change from a metal/metal Au/Ti interface (electron-electron) to a metal/non-metal Au/Ti (Au/TiO$_2$) interface (phonon-phonon) as $\Delta R$ = 6.0 m$^2$ K GW$^{-1}$ (6.6 m$^2$ K GW$^{-1}$). While this calculation of $\Delta R$ is slightly lower than our observed change in thermal boundary resistance across the Au/Ti/Gr/SiO$_2$ interfaces ($\Delta R$ ~15.9 m$^2$ K GW$^{-1}$), we caution that the assumptions required for DMM predictions could lead to uncertainties in these predicted values. Regardless, a clear change in Au/Ti thermal boundary conductance will occur when the interfacial transport transitions from an electron to phonon dominated process.

These simple qualitative and quantitative analyses suggest that the changes in thermal boundary conductance across the Au/Ti/Gr/SiO$_2$ boundary originate from changes in resistance at the Au/Ti interface and possible additional changes in thermal conductivity in the Ti layer. However, much more work needs to be pursued to study this precise interface in more detail and to understand the fundamental electron and phonon scattering mechanisms driving this thermal transport process with respect to changes in oxygen chemistry. This points to the future promise of manipulating metal/metal contacts through metal type and chemistry to impact the thermal resistances of graphene devices.

CONCLUSION

This work sheds light on the inherent variability in graphene devices. By attempting to correlate deposition conditions with the contact composition and contact resistance, we have found that contact resistance is sensitive to the partial pressure during contact deposition, and that the oxide of a Ti contact can strongly impact the thermal boundary conductance. It should be noted that reactor pressure and deposition rate are not parameters that are typically reported when describing device fabrication and yet this work demonstrates that both clearly affect device properties. The relationship between interface chemistry and contact resistance as well as thermal transport opens doors for interface engineering. While the role of interface morphology has not been explored in this study, we intend to examine it in future work.

ACKNOWLEDGMENTS

This work was supported in part by the Army Research Office, Grant No. W911NF-16-1-0320.

The Authors thank Prof. Suzanne Mohney for useful discussions. They also thank Genevieve Glista for her work on optimization of the graphene transfer process.

SUPPORTING INFORMATION

**Compositional Analysis**

X-ray photoelectron spectroscopy of the Ti 2*p* core level was used to quantify the composition of oxide and metal components. The spectra were fit using kolXPD software.[23] An example fit is shown in Figure S1 where the metal peaks are fit with a Doniach-Sunjic lineshape convoluted with a Gaussian, and the oxide peaks are fit with a Voigt lineshape. The integrated areas, or amplitudes, of the 2*p* core level peaks corresponding to metal ($I_{metal}$) and oxide ($I_{oxide}$) are used to calculate % oxide as follows:

$$\% \ oxide = \frac{I_{oxide}}{I_{oxide} + I_{metal}} \times 100$$

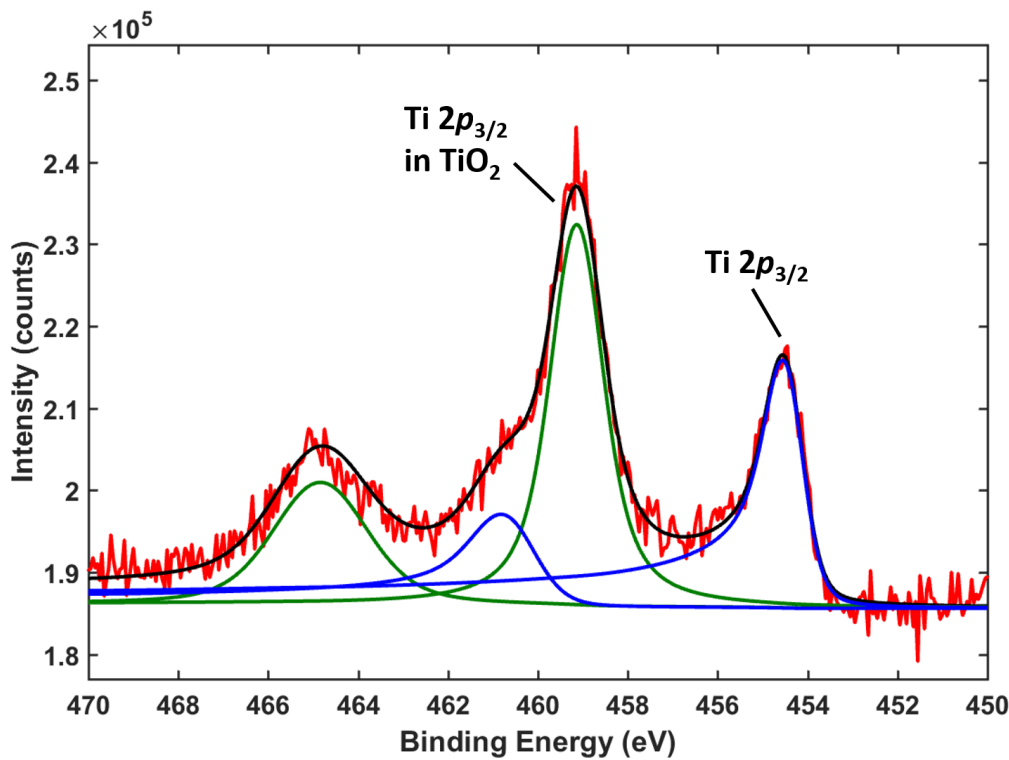

**Figure S1.** Example of peak deconvolution of a Ti 2*p* spectrum for Ti deposited on Gr/SiO$_2$ (base pressure 1x10$^{-7}$ Torr, 0.01 nm/s deposition rate for this particular sample)

**Device Structure for Transfer Length Measurement (TLM)**

The TLM structure fabricated in this work, described previously by Foley *et al*[13], is shown in Figures S2 and S3.

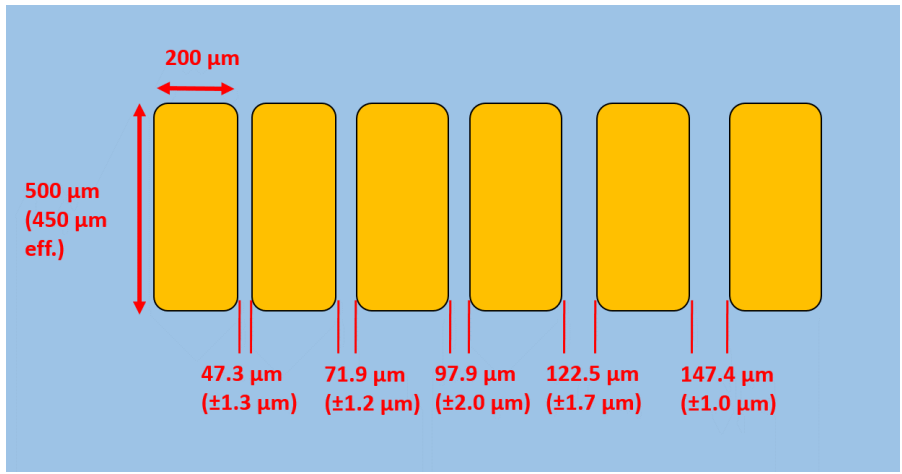

**Figure S2.** Top view of TLM structure, adapted from Ref. 13.

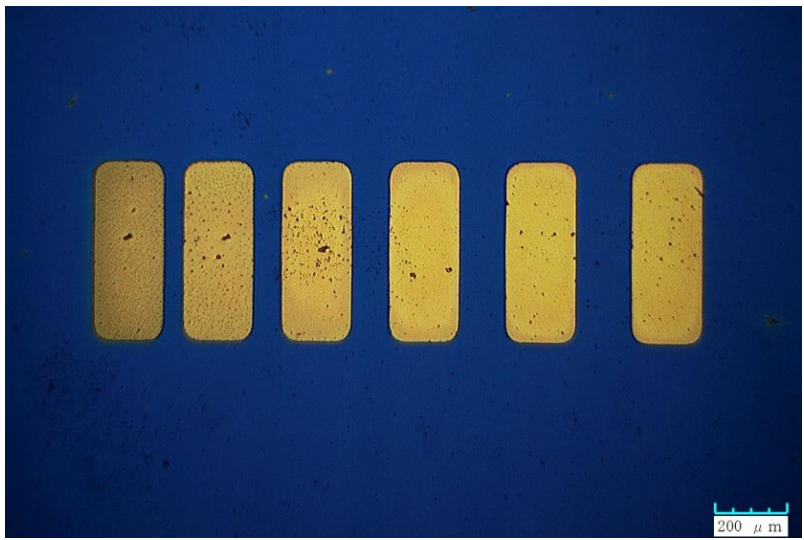

**Figure S3.** Optical micrographs of TLM device fabricated in this work.

**Coefficient of Linear Correlation for R$_C$ vs. Oxide Composition**

The extent which there exists a linear correlation between a set of points $(x_1,y_1)\ldots(x_N,y_N)$ is measured by the linear correlation coefficient, $r$, given by[35]

$$r = \frac{\sigma_{xy}}{\sigma_x \sigma_y} \qquad \text{(Eq. 1)}$$

where $\sigma_{xy}$ is the covariance, and $\sigma_x$ and $\sigma_y$ are the standard deviations of $x$ and $y$. Eq. 1 can then be written as

$$r = \frac{\sum(x_i-\bar{x})(y_i-\bar{y})}{\sqrt{\sum(x_i-\bar{x})^2 \sum(y_i-\bar{y})^2}} \qquad \text{(Eq. 2)}$$

If all points $(x_i,y_i)$ lie exactly on the line $y_i = A + Bx$ then the value of $r$ will be $\pm 1$. The quantitative significance of $r$ depends on the number of measurements, $N$, which determines the probability that two *uncorrelated* variables will yield a particular value of $r$. This can be applied conversely to determine the probability that a particular value of $r$ indicates that two variables are *correlated*. For the data reported in this work plotted in Figure 1 of the text, the measurement of oxide composition and R$_C$ on thirteen distinct samples yielded a correlation coefficient of 0.7. By the methods reported in Ref. 1, this represents a 0.8% probability that oxide composition and R$_C$ are *uncorrelated*. We therefore infer a 99.2% probability that R$_C$ is linearly correlated with oxide composition. This value corresponds to a *highly significant* probability of linear correlation.

**Time Domain Thermoreflectance**

The time domain thermoreflectance results for all four samples represented in Figure 4(a) in the text is shown here in Figure S4.

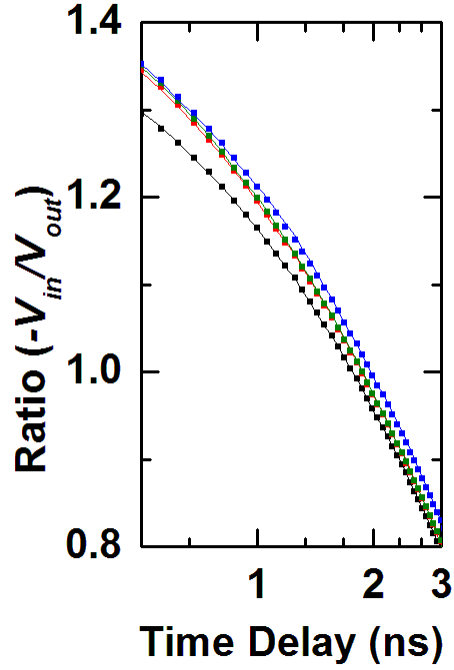

**Figure S4.** Time-domain thermoreflectance data corresponding to Gr/SiO2 samples with Ti deposited at different rates. The black curve is 0.01 nm/s, the blue curve is 0.05 nm/s the green curve is 0.1 nm/s and the red curve is 0.5 nm/s.

The respective oxide compositions and values of $h_K$ (measured by TDTR) and $R_C$ (measured by TLM) for each of these are summarized in Table S.1.

**Table S.1.** Oxide composition, thermal boundary conductance and contact resistance for different Ti dep rates

| Deposition Rate (nm/s) | Oxide Composition (%) | $h_K$ (MW m$^{-2}$ K$^{-1}$) | $R_C$ (kΩ-μm) |
|---|---|---|---|
| 0.01 | 76 | 32±3 | 4 |
| 0.05 | 16 | 39±4 | 5 |
| 0.1 | 8 | 65±7 | 5 |
| 0.5 | 10 | 58±6 | 2.6 |